\documentclass[iop]{emulateapj}
\usepackage{color}
\usepackage{savesym}
\savesymbol{tablenum}
\usepackage{amsmath,amssymb,siunitx}
\restoresymbol{aug}{tablenum}
\shorttitle{}
\shortauthors{Kyutoku \& Ioka}

\begin{document}

\begin{flushright}
 \quad \\
 \quad \\
 \quad \\
YITP-16-30\\
KEK-TH 1892\\
KEK-Cosmo 191
\end{flushright}

\title{The unreasonable weakness of \textit{r}-process cosmic rays in
the neutron-star-merger nucleosynthesis scenario}
\author{Koutarou Kyutoku\altaffilmark{1} and Kunihito
Ioka\altaffilmark{2,3,4}}
\altaffiltext{1}{Interdisciplinary Theoretical Science (iTHES) Research
Group, RIKEN, Wako, Saitama 351-0198, Japan; koutarou.kyutoku@riken.jp}
\altaffiltext{2}{Center for Gravitational Physics, Yukawa Institute for
Theoretical Physics, Kyoto University, Kyoto, 606-8502, Japan}
\altaffiltext{3}{Theory Center, Institute of Particles and Nuclear
Studies, KEK, Tsukuba 305-0801, Japan}
\altaffiltext{4}{Department of Particle and Nuclear Physics, the
Graduate University for Advanced Studies (Sokendai), Tsukuba 305-0801,
Japan}

\begin{abstract}
 We reach the robust conclusion that, by combining the observed cosmic
 rays of \textit{r}-process elements with the fact that the velocity of
 the neutron-star-merger ejecta is much higher than that of the
 supernova ejecta, either (1) the reverse shock in the
 neutron-star-merger ejecta is a very inefficient accelerator that
 converts less than 0.003\% of the ejecta kinetic energy to the
 cosmic-ray energy or (2) the neutron star merger is not the origin of
 the Galactic \textit{r}-process elements. We also find that the
 acceleration efficiency should be less than 0.1\% for the reverse shock
 of the supernova ejecta with the observed cosmic rays lighter than the
 iron.
\end{abstract}

\maketitle

\section{Introduction}

Where are the \textit{r}-process elements---neutron-rich, heavy elements
such as the gold, platinum, and rare earth elements---formed in our
universe? This is a longstanding problem in nuclear astrophysics
\citep[see][for reviews]{arnould_gt2007,qian_wasserburg2007}. The
\textit{r}-process nucleosynthesis requires an extremely neutron-rich
environment, because the neutrons have to be captured by seed nuclei
more rapidly than the $\beta$-decay proceeds. Thus, it is widely
believed that the \textit{r}-process nucleosynthesis is intimately
related to explosive events involving neutron stars.

The universality of the observed \textit{r}-process abundance pattern in
metal-poor stars suggests that only a single type of astronomical events
should be responsible for the nucleosynthesis \citep{sneden_cg2008}. One
possible site is the core-collapse supernova \citep{burbidge_bfh1957},
but theoretical investigations are gathering evidence against the
successful \textit{r}-process nucleosynthesis in this scenario
\citep{qian_woosley1996,wanajo2013} except for presumably rare
magnetorotational supernovae \citep{nishimura_tt2015}. Another possible
site is the merger of binary neutron stars and/or black hole-neutron
star binaries, which we collectively call neutron star mergers
\citep{lattimer_schramm1974,symbalisty_schramm1982}. The
neutron-star-merger scenario seems to be supported by successful
\textit{r}-process nucleosynthesis in nuclear network calculations
\citep{freiburghaus_rt1999,wanajo_snkks2014}, consistency of the
theoretically estimated production rate with the observed abundance
\citep{korobkin_raw2012}, possible detections of the macronova/kilonova
following short-hard gamma-ray bursts
\citep{berger_fc2013,tanvir_lfhhwt2013,yang_jlczhfpw2015}, reproduction
of metal-poor \textit{r}-process-enriched stars in Galactic chemical
evolution models
\citep{hirai_isfhk2015,shen_crmmg2015,vandevoort_qhkf2015}, and
indication of a ``low-rate/high-yield'' event from the deep-sea
plutonium measurement \citep{hotokezaka_pp2015} and from the observation
of ultrafaint dwarf galaxies \citep{ji_fcs2016}.

In this paper, we revisit an old idea that the composition measurement
of cosmic rays could serve as a tool to explore the origin of elements
\citep[see, e.g.,][]{arnett_schramm1973}, aiming at critically assessing
the plausibility of the neutron-star-merger scenario for the
\textit{r}-process nucleosynthesis. Because the long-term evolution of
the neutron-star-merger ejecta should resemble that of the supernova
ejecta, i.e., supernova remnants \citep{nakar_piran2011}, cosmic rays
should be accelerated at the shock associated with the blast-wave
interaction between the ejecta and interstellar medium
\citep{kyutoku_is2013,takami_ki2014}. If the ejecta from any event
contain \textit{r}-process elements, they could be accelerated when the
reverse shock sweeps into the ejecta, and the amount of accelerated
\textit{r}-process elements (hereafter, \textit{r}-process cosmic rays)
should depend on the ejecta properties such as the mass and
velocity. This suggests that the flux of \textit{r}-process cosmic rays
could enable us to distinguish the nucleosynthesis sites, namely the
supernova or the neutron star merger.

Our finding is summarized as follows. First, the flux of
\textit{r}-process cosmic rays around the energy of
\SI{1}{GeV.nucleon^{-1}} could be enhanced by a few orders of magnitude
compared to the flux expected from the supernova forward shock with the
solar composition, if the \textit{r}-process elements are synthesized in
the neutron star merger. By contrast, the \textit{r}-process cosmic-ray
flux should not be significantly enhanced if the supernova is the
nucleosynthesis site. The reason for this difference is the high
velocity of the neutron-star-merger ejecta, which can increase the total
energy of the \textit{r}-process cosmic rays for a given mass of
\textit{r}-process elements. Second, the observed cosmic rays do not
show selective enhancement of \textit{r}-process elements and are known
to be consistent with the solar abundance at the acceleration sites
\citep{ellison_dm1997,meyer_de1997}. The weak flux of \textit{r}-process
cosmic rays indicates that either (1) the particle acceleration is very
inefficient at the reverse shock in the \textit{r}-process-dominated
neutron-star-merger ejecta or (2) the neutron star merger does not
contribute to Galactic \textit{r}-process elements significantly.

\section{Important quantities}

Before starting the discussion of \textit{r}-process cosmic rays, we
list important quantities used in later estimation. The relative mass
fraction of \textit{r}-process elements in the solar neighborhood is
estimated to be \citep{qian2000}
\begin{equation}
 X_r \sim \num{e-7}.
\end{equation}
This implies that the amount of \textit{r}-process elements in our
Galaxy is $M_r \sim \num{e4} M_\odot$ and that the Galactic
\textit{r}-process element production rate is $\dot{M}_r \sim \num{e-6}
M_\odot \, \si{yr^{-1}}$. This quantity will play a key role in our
estimation to alleviate uncertainties associated with the event rate,
ejecta mass, and \textit{r}-process yield per event.

The amount of \textit{r}-process elements synthesized in a single event
under a hypothetical production scenario is estimated by dividing the
Galactic production rate, $\dot{M}_r$, by the event rate. On one hand,
the Galactic supernova rate is estimated to be $\mathcal{R}_\mathrm{SN}
\sim \SI{3e-2}{yr^{-1}}$ \citep[see, e.g.,][]{qian2000}. Thus, the
single-event yield of \textit{r}-process elements should be
$M_{r,\mathrm{SN}} = \dot{M_r} / \mathcal{R}_\mathrm{SN} \sim \num{3e-5}
M_\odot$ if the supernova is the production site. This value should be
compared with the typical mass of the supernova ejecta,
$M_\mathrm{ej,SN} \sim 3 M_\odot$ \citep[see,
e.g.,][]{truelove_mckee1999}, and therefore the fraction of
\textit{r}-process elements is
\begin{equation}
 X_{r,\mathrm{SN}} \sim \num{e-5}.
\end{equation}
On the other hand, the Galactic neutron-star-merger rate is estimated to
be $\mathcal{R}_\mathrm{NSM} \sim \SI{e-4}{yr^{-1}}$ with a large
uncertainty \citep{ligovirgo2010}. This gives us an estimate of the
single-event yield as $M_{r,\mathrm{NSM}} = \dot{M}_r /
\mathcal{R}_\mathrm{NSM} \sim 0.01 M_\odot$ for the neutron-star-merger
scenario. Assuming that all the ejecta material is synthesized to
\textit{r}-process elements
\citep{freiburghaus_rt1999,wanajo_snkks2014}, i.e.,
\begin{equation}
 X_{r,\mathrm{NSM}} \sim 1 ,
\end{equation}
this value is consistent with the typical ejecta mass $M_\mathrm{ej,NSM}
\sim 0.01 M_\odot$ obtained by general-relativistic hydrodynamical
simulations \citep{bauswein_gj2013,hotokezaka_kkosst2013} and is also
roughly sufficient to power a possible macronova/kilonova associated
with GRB 130603B
\citep{berger_fc2013,hotokezaka_ktkssw2013,tanvir_lfhhwt2013} \citep[but
see also][]{kisaka_it2015}.

It is widely accepted that Galactic cosmic rays are accelerated from the
inter/circumstellar material at the forward shock of blast waves
associated with the supernova ejecta, whereas the acceleration mechanism
is not fully understood yet. This scenario requires a fraction
\begin{equation}
 \epsilon_\mathrm{CR} \sim 0.1
\end{equation}
of the ejecta kinetic energy to be converted to the cosmic-ray
energy. We use the fiducial value, $\epsilon_\mathrm{CR} \sim 0.1$, for
both the supernova and neutron-star-merger forward shocks in this
study. It should be cautioned that the neutron-star-merger remnants
could be different from the supernova remnants in the nonspherical
geometry and in the circumburst medium that is not affected by winds or
precursors \citep{montes_rnsl2016}.

The most uncertain quantity is the fraction of the energy given to
cosmic rays at the reverse shock to that at the forward shock,
$\eta_{r/f}$. This fraction is crucial for the total energy of the
\textit{r}-process cosmic rays, because they are expected to be
accelerated not only at the forward shock but also at the reverse shock
sweeping into the ejecta. We take a provisional value of this fraction
to be
\begin{equation}
 \eta_{r/f} \sim 0.01
\end{equation}
for both the supernova and neutron-star-merger ejecta, because this is
an upper limit derived by the condition that the reverse-shock
acceleration does not dominate the forward-shock acceleration for the
elements synthesized in the supernova (see the next section). The
discussion on the \textit{r}-process cosmic rays will impose a tighter
constraint on the acceleration efficiency at the reverse shock,
$\epsilon_{\mathrm{CR},r} \equiv \eta_{r/f} \epsilon_\mathrm{CR}$, in
the neutron-star-merger scenario.

Before closing this section, we caution that both $\epsilon_\mathrm{CR}$
and $\eta_{r/f}$ can be different between the supernova ejecta and the
neutron-star-merger ejecta. In the next section, we tentatively take
common values for them to estimate \textit{r}-process cosmic rays in the
neutron-star-merger nucleosynthesis scenario. After that, we derive
constraints on these parameters separately for the supernova and the
neutron star merger.

\section{\textit{r}-process cosmic rays}

We focus on the flux at the lowest energy range for \textit{r}-process
cosmic rays, $\sim \SI{1}{GeV.nucleon^{-1}}$. In this range, the flux is
determined primarily by the total energy or injection rate of
\textit{r}-process cosmic rays irrespective of the spectral index, the
prediction of which requires the precise knowledge of acceleration,
escape, and propagation. The Galactic confinement time does not affect
comparisons among different acceleration sites, because it should be
determined solely by the rigidity. Although the composition changes
moderately during the Galactic propagation, this again does not affect
the comparisons.

First, we estimate the energy injection rate of \textit{r}-process
cosmic rays out of the inter/circumstellar material at the forward shock
associated with the supernova ejecta. Taking the typical kinetic energy
of the supernova ejecta to be $E_\mathrm{ej,SN} \sim \SI{e51}{erg}$
\citep[see, e.g.,][]{truelove_mckee1999}, the energy injection rate of
cosmic rays is estimated to be $\dot{E}_\mathrm{CR} =
\epsilon_\mathrm{CR} E_\mathrm{ej,SN} \mathcal{R}_\mathrm{SN} \sim
\SI{3e48}{erg.yr^{-1}}$. This amount roughly matches the total
generation rate of Galactic cosmic rays estimated from the observation.
If we assume that the cosmic-ray composition ratio is proportional to
the elemental abundance, the energy injection rate of \textit{r}-process
cosmic rays is given by
\begin{equation}
 \dot{E}_{r,\mathrm{ISM}} = X_r \dot{E}_\mathrm{CR} \sim
  \SI{3e41}{erg.yr^{-1}} .
\end{equation}
We consider the enhancement of refractory and/or large-mass-number
elements \citep{ellison_dm1997,meyer_de1997} separately in the next
section. The contribution from the forward shock associated with the
neutron-star-merger ejecta is negligible due to the low event rate.

Next, we estimate the energy injection rate of \textit{r}-process cosmic
rays for the supernova nucleosynthesis scenario. The energy injection
rate of all the accelerated particles at the reverse shock is given by
$\eta_{r/f} \dot{E}_\mathrm{CR} \sim \SI{3e46}{erg.yr^{-1}} ( \eta_{r/f}
/ 0.01 )$. Taking $X_{r,\mathrm{SN}} \sim \num{e-5}$, we obtain
\begin{align}
 \dot{E}_{r,\mathrm{SN}} & = X_{r,\mathrm{SN}} \eta_{r/f}
 \dot{E}_\mathrm{CR} \notag \\
 & = X_{r,\mathrm{SN}} \eta_{r/f} \epsilon_\mathrm{CR} E_\mathrm{ej,SN}
 \mathcal{R}_\mathrm{SN} \notag \\
 & \sim \SI{3e41}{erg.yr^{-1}} \left( \frac{\eta_{r/f}}{0.01} \right) ,
\end{align}
again assuming that the cosmic-ray composition is given by the relative
mass fraction. Thus, if the supernova is the \textit{r}-process
nucleosynthesis site, we do not expect significant enhancement of
\textit{r}-process cosmic rays compared to those from the
inter/circumstellar material, $\dot{E}_{r,\mathrm{ISM}}$, as far as
$\eta_{r/f} \lesssim 0.01$.

Finally, we estimate the energy injection rate of \textit{r}-process
cosmic rays for the neutron-star-merger nucleosynthesis scenario. We
take the ejecta kinetic energy to be a typical value derived by
general-relativistic simulations of binary neutron star mergers,
$E_\mathrm{ej,NSM} \sim \SI{3e50}{erg}$
\citep{bauswein_gj2013,hotokezaka_kkosst2013}, which corresponds to the
ejecta velocity $v_\mathrm{ej,NSM} \sim 0.2 c$ with $c$ as the speed of
light. Using this value, the energy injection rate of \textit{r}-process
cosmic rays is given by
\begin{align}
 \dot{E}_{r,\mathrm{NSM}} & = X_{r,\mathrm{NSM}} \eta_{r/f}
 \epsilon_\mathrm{CR} E_\mathrm{ej,NSM} \mathcal{R}_\mathrm{NSM}
 \notag \\
 & \sim \SI{3e43}{erg.yr^{-1}} \left( \frac{\eta_{r/f}}{0.01} \right) ,
\end{align}
for our fiducial $\epsilon_\mathrm{CR} \sim 0.1$.

To summarize, we obtain the ratios
\begin{equation}
 \dot{E}_{r,\mathrm{NSM}} \sim 100 \dot{E}_{r,\mathrm{SN}} \sim 100
  \left( \frac{\eta_{r/f}}{0.01} \right) \dot{E}_{r,\mathrm{ISM}}
  , \label{eq:relation}
\end{equation}
if we adopt the same values of $\epsilon_\mathrm{CR}$ and $\eta_{r/f}$
for both scenarios. This suggests that the flux of \textit{r}-process
cosmic rays at $\sim \SI{1}{GeV.nucleon^{-1}}$ could be enhanced by a
few orders of magnitude compared to the expectation from the supernova
forward shock with the solar composition if the \textit{r}-process
nucleosynthesis occurs in the neutron star merger, while the enhancement
is not expected in the supernova scenario. Note that the relation
$\dot{E}_{X,\mathrm{SN}} \sim ( \eta_{r/f} / 0.01 )
\dot{E}_{X,\mathrm{ISM}}$ also holds for any lighter-than-iron product
\textit{X} of the supernova nucleosynthesis, so that the upper limit on
the acceleration efficiency from the supernova nucleosynthesis is
$\eta_{r/f} \lesssim 0.01$. Otherwise, the cosmic-ray composition below
the iron could have changed substantially from the solar composition
(beyond the spallation modification).

The enormous enhancement for the neutron-star-merger scenario, despite
the same total \textit{r}-process yields as the supernova scenario, is
ascribed to the high ejecta velocity. To put it simply, the energy
injection rate of \textit{r}-process cosmic rays at the reverse shock is
given by
\begin{equation}
 \dot{E}_{r,*} = \eta_{r/f} \epsilon_\mathrm{CR} \dot{M}_r
  v_{\mathrm{ej},*}^2 / 2 , \label{eq:vel}
\end{equation}
where $*$ stands for either SN or NSM, and thus the energy injection
rate at the reverse shock is proportional to the squared ejecta velocity
$v_{\mathrm{ej},*}^2$, which is larger by $\sim 100$ for the neutron
star merger than for the supernova. This ratio is completely free from
uncertainties associated with $X_{r,*}$, $M_{\mathrm{ej},*}$, and
$\mathcal{R}_*$. The value of $v_\mathrm{ej,NSM}^2$ may span a range of
$0.01$--$0.1 c^2$ \citep{bauswein_gj2013,hotokezaka_kkosst2013}, and
therefore its uncertainty may be much smaller than those of the event
rate and ejecta mass.

The above discussions are applicable even if the \textit{r}-process
elements are produced by rare events like the magnetorotational
supernova \citep[e.g.,][]{nishimura_tt2015}. Given that the ejecta mass
and energy are comparable to the normal supernova, the ejecta velocity
is similar to $v_\mathrm{ej,SN}$, so that Eqs.~\eqref{eq:relation} and
\eqref{eq:vel} remain valid.

\section{Weakness thereof}

The observations of heavy cosmic rays can be explained solely by the
forward shock of supernova blast waves without invoking any contribution
from the reverse shock
\citep{binns_ggikknsw1989,westphal_pwa1998,donnelly_todddw2012}. They
find (1) similar enhancement by a factor of 20--30 for iron-group and
refractory \textit{r/s}-process cosmic rays and (2) enhancement by a
factor of 3--10 for volatile \textit{r/s}-process cosmic rays. It is
also found that the enhancement ratio may be correlated with the mass
number of elements \citep{rauch_etal2009}. It has to be cautioned that
the measured elements are only classified by the charge, and the isotope
ratios are not accurately measured. This precludes the clear distinction
between \textit{r}-process and \textit{s}-process elements. However, no
systematic difference in enhancement is found between
\textit{r}-process-dominant elements such as the platinum and
\textit{s}-process-dominant elements such as the cerium \citep[see,
e.g., Table 1 of][for the breakdown]{sneden_cg2008}, and thus the
enhancement is likely to be similar for both elements. The strong
enhancement of the refractory elements is explained by efficient
acceleration via suprathermal injection from dust grains
\citep{ellison_dm1997,meyer_de1997,rauch_etal2009}. Furthermore, the
enhancement of \textit{s}-process cosmic rays is explained by neither
the supernova nor neutron-star-merger ejecta, because the ejecta are
hardly expected to be enriched by \textit{s}-process elements.

The observations do not support Eq.~\eqref{eq:relation} with $\eta_{r/f}
\sim 0.01$, i.e., the selective \textit{r}-process enhancement in heavy
cosmic rays. This indicates the weakness of \textit{r}-process cosmic
rays from the neutron star merger. The contribution from the
neutron-star-merger ejecta to the enhancement should be less than a
factor of three, i.e., $\dot{E}_{r,\mathrm{NSM}} \lesssim 3
\dot{E}_{r,\mathrm{ISM}}$; otherwise it contradicts the observations of
refractory elements such as the zirconium relative to the iron and with
the observations of volatile elements such as the krypton
\citep{ellison_dm1997,meyer_de1997}.

\begin{figure}
 \includegraphics[width=.95\linewidth]{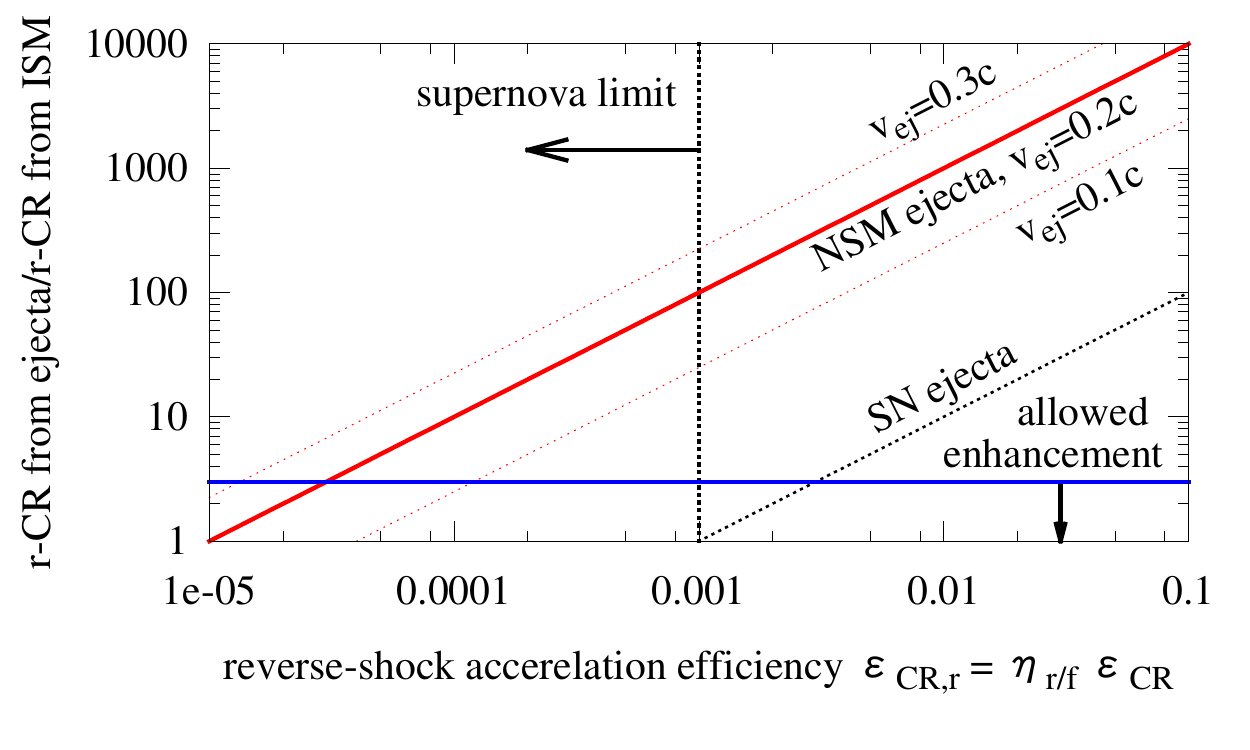} \caption{Ratio of the
 \textit{r}-process cosmic-ray flux from the reverse shock in the ejecta
 to the flux from the forward shock in the inter/circumstellar medium
 with the solar composition as a function of the acceleration efficiency
 at the reverse shock, $\epsilon_{\mathrm{CR},r} = \eta_{r/f}
 \epsilon_\mathrm{CR}$. The diagonal red lines show the enhancement
 ratio in the neutron-star-merger scenario for the \textit{r}-process
 nucleosynthesis, $\dot{E}_{r,\mathrm{NSM}} /
 \dot{E}_{r,\mathrm{ISM}}$. The solid line is for our fiducial ejecta
 velocity, $v_\mathrm{ej,NSM} = 0.2c$. The upper and lower dotted lines
 are for $0.3c$ and $0.1c$, respectively, and are drawn to indicate the
 uncertainty estimate. The diagonal black dotted line at the bottom
 right is for the supernova nucleosynthesis scenario,
 $\dot{E}_{r,\mathrm{SN}} / \dot{E}_{r,\mathrm{ISM}}$. The horizontal
 blue line is the upper limit, $\dot{E}_{r,*} / \dot{E}_{r,\mathrm{ISM}}
 \lesssim 3$, derived by the observation of heavy cosmic rays
 \citep{ellison_dm1997,meyer_de1997}. Only the region below this line is
 allowed, and thus the acceleration efficiency higher than
 $\epsilon_{\mathrm{CR},r} \sim 0.003\%$ at the reverse shock is
 unlikely in the neutron-star-merger nucleosynthesis scenario. The
 vertical dotted line is the upper limit, $\epsilon_{\mathrm{CR},r} =
 0.1\%$, obtained from the condition that the lighter-than-iron cosmic
 rays are not dominated by the supernova reverse-shock component.}
 \label{fig:param}
\end{figure}

The weakness or even absence of enhanced \textit{r}-process cosmic rays
places a constraint on particle acceleration at the reverse shock of
blast waves associated with the neutron-star-merger ejecta. Figure
\ref{fig:param} shows the allowed acceleration efficiency of
\textit{r}-process cosmic rays at the reverse shock in the
neutron-star-merger nucleosynthesis scenario. This figure indicates
that, if the neutron-star-merger scenario is true, the reverse shock
associated with the \textit{r}-process-dominated ejecta has to be an
inefficient accelerator with
\begin{equation}
 \epsilon_{\mathrm{CR},r} = \eta_{r/f} \epsilon_\mathrm{CR} \lesssim
  0.003\% .
\end{equation}
This constraint is more severe by a factor of $\sim 100/3 \sim 30$ than
that obtained from the observed cosmic rays of supernova nucleosynthesis
products, $\eta_{r/f} \lesssim 0.1\%$, and could serve as important
information for the physics of particle acceleration.

How unreasonable this low acceleration efficiency is depends on the
difference between the supernova and the neutron star merger. We could
assume that the cosmic-ray energy is distributed according to the mass
processed by the shock to argue that the low efficiency is a reasonable
outcome of the high-velocity ejecta. Assuming that the particle
acceleration becomes inefficient at the time when the blast-wave
velocity decreases to a fixed value of $v_\mathrm{tr} \sim
\SI{100}{\kilo\meter\per\second}$, at which the radiative cooling sets
in, the mass swept by the forward shock before the termination of
acceleration $M_{\mathrm{tr},*}$ is expected to be proportional to
$E_{\mathrm{ej},*}$ due to the relation $E_{\mathrm{ej},*} \propto
M_{\mathrm{tr},*} v_\mathrm{tr}^2$ in the Sedov--Taylor phase. Thus,
using $E_{\mathrm{ej},*} = M_{\mathrm{ej},*} v_{\mathrm{ej},*}^2 /2$, we
could expect $\eta_{r/f} \sim M_{\mathrm{ej},*} / M_{\mathrm{tr},*}
\propto M_{\mathrm{ej},*} / E_{\mathrm{ej},*} \propto
v_{\mathrm{ej},*}^{-2}$. This possibility naturally explains the
two-orders-of-magnitude-lower efficiency of the neutron-star-merger
scenario, and moreover, predicts $\eta_{r/f} \sim \num{e-3}$ for the
supernova and $\sim \num{e-5}$ for the neutron star merger satisfying
observational constraints. By contrast, we could also assume that the
cosmic-ray energy is distributed according to the energy processed by
the shock to argue that the low efficiency is unreasonable. In this
case, the cosmic-ray energy increases in a self-similar manner during
the Sedov--Taylor phase as $\dot{E}_\mathrm{CR} \propto 1/t$ in
accordance with the self-similar hydrodynamic evolution \citep[see
also][]{bell2015}. Thus, we could expect moderate dependence on the
ejecta velocity as $\eta_{r/f} \propto \ln t \propto \ln ( 1 /
v_{\mathrm{ej},*} )$, and this predicts that the efficiency varies at
most only by a factor of two to three between the supernova and the
neutron star merger with $\eta_{r/f} \sim 0.01$ for both cases.

It is worthwhile to consider other reasons for the inefficiency of the
reverse-shock acceleration in the neutron star merger. They may include
weak magnetic fields in the ejecta, absence of \textit{r}-process dust
grains, and/or energy loss to the adiabatic expansion. The absence of
\textit{r}-process dust grains in the neutron-star-merger ejecta is
consistent with the finding of \citet{takami_ni2014}, although this
effect is not sufficient to fully account for the weakness. Conversely,
if the reverse shock turns out to be a moderately efficient accelerator
for the neutron-star-merger ejecta, our result implies that the neutron
star merger contributes very little or not at all to the Galactic
\textit{r}-process enrichment.

\section{Future prospect}

Precise measurement of the cosmic-ray composition is useful to narrow
down the uncertainty range. In particular, a detailed investigation of
the difference between \textit{r}-process and \textit{s}-process
cosmic-ray fluxes is highly appreciated, because the contribution from
any ejecta is likely to be much larger in the \textit{r}-process
enhancement than in the \textit{s}-process enhancement. Moreover,
uncertainties should be reduced if we compare elements with similar mass
numbers and similar volatility. For this purpose, it is important to
measure the isotope ratio of heavy cosmic rays to precisely separate
\textit{r}-process and \textit{s}-process elements. We also have to
understand nuclear interactions like spallation during the cosmic-ray
propagation to precisely recover the composition at the acceleration
site from the observed one.

The velocity of the neutron-star-merger ejecta also requires an
investigation. While it is safely expected that the ejecta velocity is
higher for the neutron star merger than for the supernova, the precise
value depends on the mass ejection mechanism. For example, the velocity
is likely to become low if the late-time activity such as disk winds
contributes substantially to the mass ejection
\citep{fernandez_metzger2013,just_bagj2015,kiuchi_skstw2015}. The
velocity could be determined by observing synchrotron emission around
the Sedov time \citep{nakar_piran2011,takami_ki2014}.

It is very difficult to pin down the acceleration efficiency at the
reverse shock accurately, while this is one of the last pieces to assess
the plausibility of the neutron-star-merger scenario for the
\textit{r}-process nucleosynthesis by studying cosmic rays. We could in
principle infer the efficiency by observing reverse-shock emission, such
as gamma-rays from hadronic interaction \citep[see, e.g.,][for leptonic
emission from Cas A]{helder_vink2008}. However, the large distance of
$\sim \SI{100}{Mpc}$ expected for a yearly neutron-star-merger event
prohibits us from taking this approach in the real life
\citep{ligovirgo2010}.

Theoretical investigations of the reverse-shock acceleration are highly
desired, particularly for the \textit{r}-process-dominated
ejecta. Relevant topics include magnetic-field amplification from the
value typical of neutron stars, injection of \textit{r}-process elements
involving possible dust formation, and energy loss to the adiabatic
expansion before the escape. If efficient acceleration at the reverse
shock would be a likely outcome, the unreasonable weakness of
\textit{r}-process cosmic rays challenges the neutron-star-merger
nucleosynthesis scenario.

\begin{acknowledgments}
 We are deeply indebted to Hajime Takami for the help during the early
 stage of this work, and thank Sho Fujibayashi, Yutaka Ohira, Shinya
 Wanajo, and Ryo Yamazaki for valuable discussions. We also thank the
 anonymous referee for carrying the burden of critical review. This work
 is supported by JSPS KAKENHI Grant-in-Aid for Specially promoted
 Research (No.~24000004), for Scientific Research on Innovative Areas
 (No.~24103006), for Scientific Research (No.~26247042, No.~26287051),
 and for Research Activity Start-up (No.~15H06857). K.K. is supported by
 the RIKEN iTHES project.
\end{acknowledgments}

\end{document}